# Vulnerable Road User Detection Using Smartphone Sensors and Recurrence Quantification Analysis

Huthaifa I. Ashqar, Mohammed Elhenawy, Mahmoud Masoud, Andry Rakotonirainy, and Hesham A. Rakha, *IEEE, Senior Member*

*Abstract*—With the fast advancements of the Autonomous Vehicle (AV) industry, detection of Vulnerable Road Users (VRUs) using smartphones is critical for safety applications of Cooperative Intelligent Transportation Systems (C-ITSs). This study explores the use of low-power smartphone sensors and the Recurrence Quantification Analysis (RQA) features for this task. These features are computed over a thresholded similarity matrix extracted from nine channels: accelerometer, gyroscope, and rotation vector in each direction ($x, y,$ and $z$). Given the high-power consumption of GPS, GPS data is excluded. RQA features are added to traditional time domain features to investigate the classification accuracy when using binary, four-class, and five-class Random Forest classifiers. Experimental results show a promising performance when only using RQA features with a resulted accuracy of $98.34\%$ and a $98.79\%$ by adding time domain features. Results outperform previous reported accuracy, demonstrating that RQA features have high classifying capability with respect to VRU detection.

I. INTRODUCTION

With the continuous development of the Autonomous Vehicle (AV) industry, the detection of Vulnerable Road Users (VRUs) (e.g. cyclists, runners, and pedestrians) is becoming critical in building Cooperative Intelligent Transportation System (C-ITS) safety applications to improve VRU safety [1, 2]. As the first generation ITS is unable to share data and cooperate, C-ITS was introduced and is currently being researched. C-ITS aims to improve safety, sustainability, efficiency and comfort using different communication technologies including vehicle-to-vehicle (V2V), vehicle-to-infrastructure (V2I), infrastructure-to-vehicle (I2V) and infrastructure-to-infrastructure (I2I). C-ITS exchanges several types of information such as information regarding Vulnerable Road Users (VRUs), traffic jams, accidents, and road hazards [3]. This enables C-ITS to build an integrated system of people, roads, infrastructure, and vehicles by applying communications, computers, and other technologies in the field of transportation. Integrating information and technologies can establish a large, full-functioning, real-time, accurate, efficient, and safe transportation system [2, 4]. Although C-ITS is currently under the spotlight, researchers mainly focused on non-VRUs such as cars, buses, and trucks, showing less concern for VRUs [5]. One of the important key planning issues facing deploying AV is the inter-modal traffic regulation, in which AV policies and programming should be designed to respect human life by minimizing crash risks and protect VRUs (e.g., through lower speeds) [6, 7].

VRUs are considered one example of the bystander humans interfering with AVs, which could be "*humans that do not explicitly interact with the automated vehicle but still affect how the vehicle accomplishes its task by observing or interfering with the actions of the vehicle.*"[1] VRUs, which do not usually have a protective 'shell', can be defined as 'vulnerable' in a number of ways. In this study, we defined them by the amount of protection in traffic (i.e. pedestrians, runners, and cyclists). VRU lethality of pedestrians and motorcyclists is higher than average as the inequality factor for collisions between VRUs is low ($< 10$) [8].

On the other hand, the application of smartphones to data collection has also recently attracted researchers' attention. Smartphone applications (apps) have been developed and effectively used to collect data from smartphones in many sectors. In the transportation sector, researchers can use smartphones to track and obtain information such as speed, acceleration, and the rotation vector from the built-in Global Positioning System (GPS), accelerometer, and gyroscope sensors. These data can be used to recognize the user's transportation mode, which can then be utilized in a number of different applications and could significantly reduce the time and cost for conventional travel surveys, as shown in Table I.

In this study, we investigate the possibility of detecting VRUs with high accuracy using data collected from smartphone low-power sensors. Moreover, we explore the use of RQA features and pooling them with the traditionally-used time domain features to boost the classification accuracy.

H. I. Ashqar is with Booz Allen Hamilton, Washington, D.C., 20003 USA (e-mail: hiashqar@vt.edu).

M. Elhenawy is with the Centre for Accident Research and Road Safety, Queensland University of Technology, Kelvin Grove QLD 4059 Australia (e-mail: mohammed.elhenawy@qut.edu.au).

M. Masoud is with the Centre for Accident Research and Road Safety, Queensland University of Technology, Kelvin Grove QLD 4059 Australia (e-mail: mahmoud.masoud@qut.edu.au).

A. Rakotonirainy is with the Centre for Accident Research and Road Safety, Queensland University of Technology, Kelvin Grove QLD 4059 Australia (e-mail: r.andry@qut.edu.au).

H. A. Rakha is with the Charles E. Via, Jr. Department of Civil and Environmental Engineering, Virginia Tech, VA 24061 USA (e-mail: hrakha@vt.edu).

TABLE I. TRANSPORTATION MODE DETECTION APPLICATIONS [9].

| Application | Description |
|---|---|
| *Autonomous Vehicle (AV)* | Development of a cooperative system for the real-time detection and relative localization of VRU using C-ITS technologies to improve safety and provide protection for VRU. Moreover, VRU detection provides broad understanding for the problem of human trust in autonomous vehicles [1]. |
| *Transportation Planning* | Instead of using traditional approaches such as questionnaires, travel diaries, and telephone interviews [10, 11], the transportation mode information can be automatically obtained through smartphone sensors. |
| *Safety* | Knowing the transportation mode can help in developing safety applications. For example, violation prediction models have been studied for passenger cars and bicycles [12]. |
| *Environment* | Physical activities, health, and calories burned, and carbon footprint associated with each transportation mode can be obtained when the mode information is available [13]. |
| *Information Provision* | Traveler information can be provided based on the transportation mode [11, 14]. |

## II. RELATED WORK

Researchers have developed several approaches to discriminate between transportation modes effectively. Machine learning techniques have been used extensively to build detection models and have shown high accuracy in determining transportation modes. Supervised learning methods such as K-Nearest Neighbor (KNN) [15], Support Vector Machines (SVMs) [16-18], Decision Trees [10, 11, 13, 14, 18-20], and Random Forests (RFs) [15, 18], have all been employed in various studies.

These studies have obtained different classifying accuracies. There are several factors that affect the accuracy of detecting transportation modes, such as the *monitoring period* (positive association), *number of modes* (negative association), *data sources, motorized classes,* and *sensor positioning* [9, 15]. However, one of the most critical factors that affects the accuracy of mode detection is the machine learning framework classifier. The framework that usually uses one layer of classification algorithm as in [11, 13, 15] could be refered to as a traditional framework; whereas the hierarchal framework uses more than one layer of classification algorithms as in [18].

An additional important consideration is the domain of the extracted features. Features are generally extracted from two different domains: (1) the time domain, features of which have been used widely in many studies [11, 13, 15-18, 21]; and (2) the frequency domain, features of which have been used in some studies [13, 16, 18]. Both methods have achieved a significant, high accuracy. Table II summarizes the obtained accuracies and factors for some of the aforementioned studies. Note that no direct comparison can be made between the studies listed in Table II because the factors considered, and the data sets used varied from study to study.

Specifically, in terms of VRU detection, Zadeh *et al.* [22] developed a new warning system on smartphones to protect VRUs (i.e. children and elderly). The system proposes a geometric approach to activate the system for necessary risky situations: low, medium, and high risk. By extracting features from smartphone sensors for VRUs and drivers, the system estimates the collision risk by using a fuzzy inference engine. Moreover, Fardi *et al.* [23] used image processing to provide pedestrian recognition using a 3D photonic mixer device (PMD) camera. The presented sensor system was able to fulfil specific requirements for pedestrian protection assistant. Furthermore, Anaya *et al.* [5] developed a novel Advanced Driver Assistance System (ADAS) to avoid incidents that involve motorcyclist and cyclists using V2V communications. However, Gavrila *et al.* [24] described a multi-sensor approach for the protection of VRUs as a part of the PROTECTOR project (Preventive Safety for Unprotected Road User) by detecting VRUs and distinguishing them from a moving vehicle.

This study mainly investigates the effect of extracted features on the accuracy of VRU detection by exploring the use of RQA analysis. The main goal is to accurately detect VRUs using data collected from the smartphone low-power sensors. High level C-ITS safety functions depend on an accurate VRU detection. Based on different applications, three classifiers are introduced using RF. First, a binary classifier to discriminate VRUs from non-VRUs. This binary classifier is useful in cases where a VRU is exposed to higher risks. For example, at an intersection, the smartphones of all subjects detect VRU reporting to the road side unit of the C-ITS. C-ITS receives the messages from the vehicle unit then broadcasts an appropriate warning message if it predicts a possible conflict between a VRU and the vehicle. Second, a four-class classifier, which identifies the exact mode of the VRU and identifies all non-VRU modes as a one mode. This classifier is useful for some C-ITS applications such as protecting bikes in a road by ensuring a safe passing distance and speed between vehicles and the cyclist. Third, a five-class classifier to detect all modes, which could be used for many applications.

We should highlight that most of the proposed methods in the most recent studies rely on the usage of GPS data, which do not take into account the limitations of GPS information. GPS service is not available or may be lost in some areas, which results in inaccurate position information. Moreover, the GPS system use might deplete the smartphone's battery. Thus, this study focuses on using data obtained from different smartphone sensors excluding GPS data.

TABLE II. SUMMARY OF SOME PAST STUDIES [9].

| Accuracy (%) | Features Domain | Machine Learning Framework | Monitoring Period | No. of Modes | Data Sources | More than One Motorized Mode? | Sensor Positioning | Data Set | Study |
|---|---|---|---|---|---|---|---|---|---|
| 97.31 | Time | Traditional | 4 s | 3 | Accelerometer | Yes | No requirements | Not mentioned | [17] |
| 93.88 | Frequency | Traditional | 5 s, 50% overlap | 6 | Accelerometer | Yes/No | Participants were asked to keep their device in the pocket of their non-dominant hip | Collected from 4 participants | [16] |
| 93.60 | Time and frequency | Traditional | 1 s | 5 | Accelerometer GPS | No | No requirements | Collected from 16 participants | [13] |
| 93.50 | Time | Traditional | 30 s | 6 | GPS, GIS[a] maps | Yes | No requirements | Collected from 6 participants | [11] |
| 95.10 | Time | Traditional | 1 s | 5 | Accelerometer, gyroscope, rotation vector | Yes | No requirements | Collected from 10 participants | [15] |
| 91.60 | Time | Traditional | Entire trip | 11 | GPS, GIS maps | Yes | No requirements | Two different data sets, one of which included 1,000 participants | [21] |
| 97.02 | Time and frequency | Hierarchical | 1 s | 5 | Accelerometer, gyroscope, rotation vector | Yes | No requirements | Collected from 10 participants | [18] |

[a] GIS: Geographic Information System.

## III. DATA SET

### A. Data Collection

The data set used is available at the Virginia Tech Transportation Institute (VTTI) and was collected by Jahangiri and Rakha [15] using a smartphone app (two devices were used: A Galaxy Nexus and a Nexus 4). The app was provided to 10 travelers who work at VTTI to collect data for five different modes: driving a passenger car, bicycling, taking a bus, running, and walking. The data were collected from GPS, accelerometer, gyroscope, and rotation vector sensors and stored on the devices at the application's highest possible frequency. Data collection was conducted on different workdays (Monday through Friday) and during working hours (8:00 a.m. to 6:00 p.m.). Several factors were considered to collect realistic data reflecting natural behaviors. No specific requirement was applied in terms of sensor positioning other than carrying the smartphone in different positions that they normally do, to ensure that the data collection is less dependent on the sensor positioning. The data were collected on different road types with different speed limits in Blacksburg, Virginia, and some epochs may reflect traffic jam conditions occurring in real-world conditions. The collection of thirty minutes of data over the course of the study for each mode per person was considered sufficient.

For the purpose of comparing data with data from previous studies [15, 18], the extracted features were considered to have a meaningful relationship with different transportation modes. Furthermore, features that might be extracted from the absolute values of the rotation vector sensor were excluded. Additionally, in order to allow this framework to be implemented in cases where no GPS data were available, features that might be extracted from GPS data were also excluded.

### B. Time Domain Features

Jahangiri and Rakha [15] collected readings from the mobile sensors at a frequency of almost 25 Hz. Because the output samples of the sensors were not synchronized, the authors implemented a linear interpolation to build continuous signals from the discrete samples. Consequently, they sampled the constructed sensor signals at 100 Hz and divided the output of each sensor in each direction ($x, y$, and $z$) into non-overlapping windows ($t$) of 1-s width. From window $t$, time domain features were created by applying the measures in Table III. These measures were applied using the measurements of the data array $x_i^t$ and its derivative $\dot{x}_i^t$ for the $i^{th}$ feature from time window $t$. This resulted in 126 time domain features: all the 14 measures presented in Table III were applied to accelerometer, gyroscope, and rotation vector sensor values. As a result, the total number of features reached $14(3) + 14(3) + 14(3) = 126$ features.

TABLE III. MEASUREMENTS OF TIME DOMAIN FEATURES.

| No. | Measure | No. | Measure |
|---|---|---|---|
| 1 | $mean(x_i^t)$ | 8 | $mean(\dot{x}_i^t)$ |
| 2 | $max(x_i^t)$ | 9 | $max(\dot{x}_i^t)$ |
| 3 | $min(x_i^t)$ | 10 | $min(\dot{x}_i^t)$ |
| 4 | $variance(x_i^t)$ | 11 | $variance(\dot{x}_i^t)$ |
| 5 | $standard\ deviation(x_i^t)$ | 12 | $standard\ deviation(\dot{x}_i^t)$ |
| 6 | $range(x_i^t)$ | 13 | $range(\dot{x}_i^t)$ |
| 7 | $Interquartile\ range(x_i^t)$ | 14 | $Interquartile\ range(\dot{x}_i^t)$ |

## IV. METHODS

### A. Recurrence Quantification Analysis (RQA)

A traditional practice in transportation mode recognition is extracting features from the time domain that can be input to state-of-the-art algorithms for classification. The typical approach consists of averaging the features and using the statistics (i.e. mean, standard deviation, …, etc.), a process that destroys very important information about the temporal evolution and distribution of the features. The development of features that describe the temporal evolution to increase the classification accuracy is still not well studied in transportation mode recognition. In this study, we used Recurrence Quantification Analysis (RQA) to provide additional information on temporal dynamics. RQA is a nonlinear data analysis method used to investigate dynamical systems that quantifies the number and duration of recurrences of a dynamical system presented by its phase space trajectory. It was introduced by Eckmann *et al.* [25] as a graphical method to locate hidden recurring patterns, nonstationary, and structural changes. RQA was shown to be a powerful tool for the dynamical system analysis, which can provide a quantitative characterization of the complexity and randomness of nonlinear, nonstationary, and short signals [26-32]. RQA can lead to effective features, which are more sensitive to the variations and less sensitive to noise or other extraneous variations [30, 31].

### B. Random Forest (RF)

Breiman proposed RF as a new classification and regression technique in supervised learning [33]. The RF method randomly constructs a collection of decision trees in which each tree chooses a subset of features to grow, and the results are then obtained based on the majority votes from all trees. The number of decision trees and the selected features for each tree are user-defined parameters. The reason for choosing only a subset of features for each tree is to prevent the trees from being correlated. RF was applied in this study as a classification algorithm since it offers several advantages. For example, it runs efficiently on large datasets, can handle many input features without creating extra dummy variables, and it ranks each feature's individual contribution in the model [33, 34].

## V. RESULTS

### A. RQA Features

The first step in extracting RQA features is to quantify patterns that emerge in Recurrence Plots (RPs). RQA starts from performing a one-dimensional time series analysis that are assumed to result from a process involving several variables. Using Taken's theorem, multidimensionality can be recovered when delaying the time series and embedding it in a phase space. The distance matrix of the series is then computed and then a certain radius $r$ is defined. The radius is the maximum Euclidean distance at and below which the recurrent points are defined and displayed. Each distance below the radius $r$ is considered a recurrence pair and the distance matrix is transformed into a recurrence plot by darkening all the recurrent points. From this idea, several metrics have been developed that quantify the amount and length of lines of contiguous points in the matrix. RQA features were developed by Zbilut and Webber [35] and a detailed description can be found in [29, 30].

The extraction of RQA features requires the setting of three important parameters namely; delay (i.e. lag), dimension of the phase space, and threshold parameter $T$. Delay is chosen as the value at which the average mutual information (AMI) function is minimum. To calibrate the delay parameter for each channel, we averaged the mutual average information function over all participants and modes as shown in Figure 1 (a). The dimension of the phase space is identified using the false nearest neighbor (FNN) test as shown in Figure 1 (b).

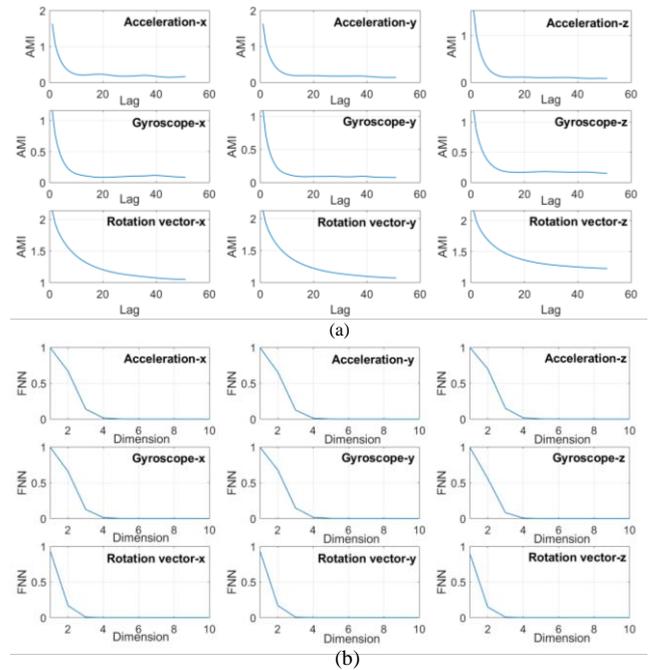

Figure 1. Results of (a) AMI function and (b) FNN test.

To determine the value of $T$, the dimension of the space and the delay were used to build the Recurrence Plot (RP) and extract RQA features at different $T$ values. The resulting RQA features from each channel were used as input in the RF algorithm. Based on the classification accuracy, $T$ was determined for each channel. Measurements of the delay, dimension of phase space, and $T$ were $10, 4, 0.9$ for accelerometer and gyroscope respectively; and were $30, 3, 0.01$ for rotation vector respectively.

Following is a summarization of the used features that resulted from the aforementioned three parameters. The extracted RQA features are: Recurrence Rate (RR), which is the percentage of points in the threshold plot; Determinism (DET), which is measured as the percentage of points that are along the diagonal lines in RP; Longest Diagonal Line (Lmax), which characterizes repetitions and is related to their periods; Entropy (ENT), which is the Shannon entropy of the diagonal line lengths; laminarity (LAM), which is the percentage of points that form vertical lines and could be useful to identify signals with stationary segments; and trapping time (TT), which is the average vertical line length that characterizes durations of stationary periods. As a result, the total number of RQA features is $6 \times 3 + 6 \times 3 + 6 \times 3 = 54$ features (i.e., 180 time and RQA features pooled in total for each 1-s window).

*B. Classification*

In this study, we used the minimum Redundancy Maximum Relevance (mRMR) [36] approach to select the most representative features of the 180 features. When selecting a feature from the feature set *A*, assuming set *B* has the already selected features, mRMR is used to simultaneously maximize the relevance between the feature and the target class and to minimize the redundancy between that feature and the already selected features. We ran a 100-tree RF with different number of features starting from the most representative features as defined by mRMR.

RF is trained and tested using data obtained from the smartphone sensors including accelerometer, gyroscope, and rotation vector sensors using cross validation. RF was used as a (1) five-class classifier (i.e. bike, walk, run, bus, and car), (2) four-class classifier (i.e. bike, walk, run, and non-VRU), and (3) binary classifier (i.e. VRU, and non-VRU). The three classifiers were all trained using pooled features from RQA and the time domain. However, the four-class and binary classifiers were then used to investigate the accuracy using RQA features only. For this purpose, we extracted 450 of RQA features from all channels.

As Figure 2(a) shows using 170 pooled features, the classification accuracy reaches 95.51%, 97.12%, and 98.79% for the five-class, four-class, and binary classifiers, respectively. These results outperform the reported results in recent studies using only time domain features. The 170 features include all RQA features and the rest are the time domain features, which means RQA features are more significant than the time domain features. As Figure 2(b) shows using 450 of RQA features, the classification accuracy reaches 90.32%, and 98.34% for the four-class and binary classifiers, respectively. RQA features have a significant discrimination power between the VRU and non-VRU (i.e. binary classifier). However, the accuracy using RQA features to classify different VRU modes is relatively low.

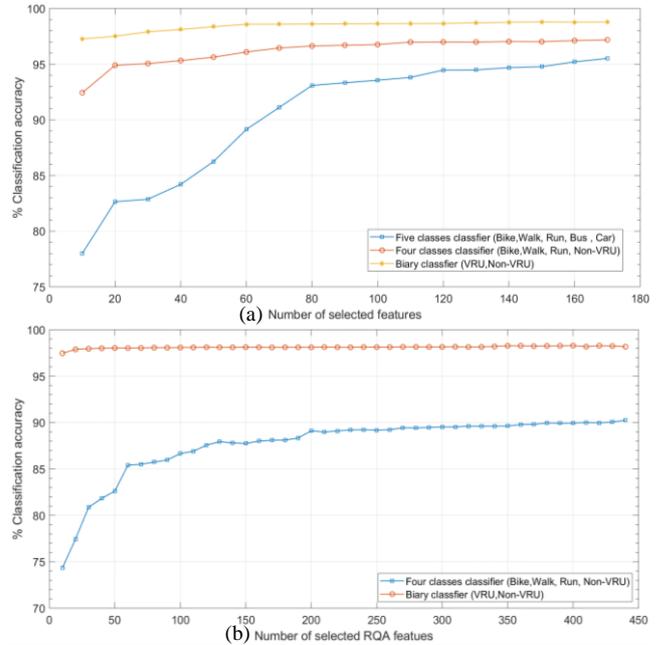

Figure 2. Accuracy of using (a) pooled features, (b) RQA features only.

Results also show that the binary classifier does not require a large number of features to achieve a very high accuracy compared to the four-class and five-class classifiers. In fact, it reaches the maximum accuracy using 60 pooled features and 10 RQA features only.

VI. CONCLUSION

With the rapid advancements of the AV industry, detection of VRUs is becoming essential to improve their safety. In this study, we investigated the effect of extracted features on the accuracy of VRU detection by exploring the use of RQA analysis that captures the temporal evolution. We extracted new features based on data obtained from smartphone sensors including accelerometer, gyroscope, and rotation vector, without GPS data given its potential limitations. The classification accuracy reaches 98.79% and 98.34% using RF with pooled features and only RQA features, respectively. Results outperform previous reported accuracy using only time domain features. Our experimental result shows that RQA features have decisive discriminating power with respect to VRU detection, that is not captured in traditional time domain features. Unlike most approaches, RQA features make no assumptions about linearity or stationarity of the data. Our next step will be including the GPS data in the classification and investigate its effect using different methods such as transfer learning.

ACKNOWLEDGMENT

This work was supported by the iMOVE Cooperative Research Centre (CRC) under Grant number 1-002.

## REFERENCES

[1] K. Saleh, M. Hossny, and S. Nahavandi, "Towards trusted autonomous vehicles from vulnerable road users perspective," in *2017 Annual IEEE International Systems Conference (SysCon)*, 2017, pp. 1-7.

[2] W. Höfs *et al.*, "International deployment of cooperative intelligent transportation systems: bilateral efforts of the European Commission and United States Department of Transportation," U.S. Department of Transportation. Research and Innovative Technology Administration (RITA), 2012.

[3] M. Alam, J. Ferreira, and J. Fonseca, "Introduction to intelligent transportation systems," pp. 1-17: Springer.

[4] M. Elhenawy, A. Bond, and A. Rakotonirainy, "C-ITS Safety Evaluation Methodology based on Cooperative Awareness Messages," in *21st International Conference on Intelligent Transportation Systems (ITSC)*, 2018, pp. 2471-2477: IEEE.

[5] J. J. Anaya, E. Talavera, D. Giménez, N. Gómez, J. Felipe, and J. E. Naranjo, "Vulnerable road users detection using v2x communications," in *IEEE 18th International Conference on Intelligent Transportation Systems*, 2015, pp. 107-112: IEEE.

[6] T. Litman, *Autonomous vehicle implementation predictions*. Victoria Transport Policy Institute Victoria, Canada, 2017.

[7] E. Papa and A. Ferreira, "Sustainable accessibility and the implementation of automated vehicles: Identifying critical decisions," *Urban Science,* vol. 2, no. 1, p. 5, 2018.

[8] SWOV, "SWOV Fact sheet: Vulnerable road users," SWOV Institute for Road Safety Research, Leidschendam, the Netherlands, 2012.

[9] M. Elhenawy, A. Jahangiri, and H. A. Rakha, "Smartphone Transportation Mode Recognition using a Hierarchical Machine Learning Classifier," presented at the 23rd ITS World Congress, Melbourne, Australia, 2016.

[10] X. Yu *et al.*, "Transportation activity analysis using smartphones," in *Consumer Communications and Networking Conference (CCNC), 2012 IEEE*, 2012, pp. 60-61.

[11] L. Stenneth, O. Wolfson, P. S. Yu, and B. Xu, "Transportation mode detection using mobile phones and GIS information," in *19th ACM SIGSPATIAL International Conference on Advances in Geographic Information Systems*, United states, 2011, pp. 54-63.

[12] A. Jahangiri, H. A. Rakha, and T. A. Dingus, "Adopting Machine Learning Methods to Predict Red-light Running Violations," in *Intelligent Transportation Systems (ITSC), 2015 IEEE 18th International Conference on*, 2015, pp. 650-655: IEEE.

[13] S. Reddy, M. Mun, J. Burke, D. Estrin, M. Hansen, and M. Srivastava, "Using mobile phones to determine transportation modes," *ACM Transactions on Sensor Networks (TOSN),* vol. 6, no. 2, p. 13, 2010.

[14] V. Manzoni, D. Maniloff, K. Kloeckl, and C. Ratti, "Transportation mode identification and real-time CO2 emission estimation using smartphones," Technical report, Massachusetts Institute of Technology, Cambridge2010.

[15] A. Jahangiri and H. A. Rakha, "Applying machine learning techniques to transportation mode recognition using mobile phone sensor data," *IEEE transactions on intelligent transportation systems,* vol. 16, no. 5, pp. 2406-2417, 2015.

[16] B. Nham, K. Siangliulue, and S. Yeung, "Predicting mode of transport from iphone accelerometer data," Tech. report, Stanford Univ2008.

[17] T. Nick, E. Coersmeier, J. Geldmacher, and J. Goetze, "Classifying means of transportation using mobile sensor data," in *Neural Networks (IJCNN), The 2010 International Joint Conference on*, 2010, pp. 1-6: IEEE.

[18] H. I. Ashqar, M. H. Almannaa, M. Elhenawy, H. A. Rakha, and L. House, "Smartphone Transportation Mode Recognition Using a Hierarchical Machine Learning Classifier and Pooled Features From Time and Frequency Domains," *IEEE Transactions on Intelligent Transportation Systems,* no. 99, pp. 1-9, 2018.

[19] Y. Zheng, L. Liu, L. Wang, and X. Xie, "Learning transportation mode from raw gps data for geographic applications on the web," presented at the Proceedings of the 17th international conference on World Wide Web, Beijing, China, 2008.

[20] P. Widhalm, P. Nitsche, and N. Brandie, "Transport mode detection with realistic Smartphone sensor data," in *2012 21st International Conference on Pattern Recognition (ICPR 2012), 11-15 Nov. 2012*, Piscataway, NJ, USA, 2012, pp. 573-6: IEEE.

[21] F. Biljecki, H. Ledoux, and P. Van Oosterom, "Transportation mode-based segmentation and classification of movement trajectories," *International Journal of Geographical Information Science,* vol. 27, no. 2, pp. 385-407, 2013.

[22] R. B. Zadeh, M. Ghatee, and H. R. Eftekhari, "Three-phases smartphone-based warning system to protect vulnerable road users under fuzzy conditions," *IEEE Transactions on Intelligent Transportation Systems,* vol. 19, no. 7, pp. 2086-2098, 2018.

[23] B. Fardi, J. Dousa, G. Wanielik, B. Elias, and A. Barke, "Obstacle detection and pedestrian recognition using a 3D PMD camera," pp. 225-230: IEEE.

[24] D. M. Gavrila, M. Kunert, and U. Lages, "A multi-sensor approach for the protection of vulnerable traffic participants the PROTECTOR project," in *Proceedings of the 18th IEEE Instrumentation and Measurement Technology Conference*, 2001, vol. 3, pp. 2044-2048: IEEE.

[25] J. Eckmann, S. O. Kamphorst, and D. Ruelle, "Recurrence plots of dynamical systems," *World Scientific Series on Nonlinear Science Series A,* vol. 16, pp. 441-446, 1995.

[26] L. Librizzi, F. Noè, A. Vezzani, M. De Curtis, and T. Ravizza, "Seizure-induced brain-borne inflammation sustains seizure recurrence and blood–brain barrier damage," *Annals of neurology,* vol. 72, no. 1, pp. 82-90, 2012.

[27] G. Roma, W. Nogueira, P. Herrera, and R. de Boronat, "Recurrence quantification analysis features for auditory scene classification," *IEEE AASP challenge on detection and classification of acoustic scenes and events,* vol. 2, 2013.

[28] L. Silva, J. R. Vaz, M. A. Castro, P. Serranho, J. Cabri, and P. Pezarat-Correia, "Recurrence quantification analysis and support vector machines for golf handicap and low back pain EMG classification," *Journal of Electromyography and Kinesiology,* vol. 25, no. 4, pp. 637-647, 2015.

[29] C. L. Webber Jr and N. Marwan, "Recurrence quantification analysis," *Theory and Best Practices,* 2015.

[30] H. Yang, "Multiscale recurrence quantification analysis of spatial cardiac vectorcardiogram signals," *IEEE Transactions on Biomedical Engineering,* vol. 58, no. 2, pp. 339-347, 2011.

[31] J. P. Zbilut, N. Thomasson, and C. L. Webber, "Recurrence quantification analysis as a tool for nonlinear exploration of nonstationary cardiac signals," *Medical engineering & physics,* vol. 24, no. 1, pp. 53-60, 2002.

[32] U. R. Acharya, S. V. Sree, G. Swapna, R. J. Martis, and J. S. Suri, "Automated EEG analysis of epilepsy: a review," *Knowledge-Based Systems,* vol. 45, pp. 147-165, 2013.

[33] L. Breiman, "Random forests," *Machine learning,* vol. 45, no. 1, pp. 5-32, 2001.

[34] W. Y. Loh, "Classification and regression trees," *Wiley Interdisciplinary Reviews: Data Mining and Knowledge Discovery,* vol. 1, no. 1, pp. 14-23, 2011.

[35] C. L. Webber Jr and J. P. Zbilut, "Dynamical assessment of physiological systems and states using recurrence plot strategies," *Journal of applied physiology,* vol. 76, no. 2, pp. 965-973, 1994.

[36] H. Peng, F. Long, and C. Ding, "Feature selection based on mutual information: criteria of max-dependency, max-relevance, and min-redundancy," *IEEE Transactions on Pattern Analysis & Machine Intelligence,* no. 8, pp. 1226-1238, 2005.